\newcommand{\br}{{\bf r}}
\newcommand{\rt}{r_\bot}
\newcommand{\brt}{{\bf r}_\bot}
\newcommand{\lt}{{\cal L}_\bot}
\newcommand{\lz}{{\cal L}_z}
\documentstyle[aps,prl,twocolumn,epsfig,graphics]{revtex}

\begin{document}
\twocolumn[\hsize\textwidth\columnwidth\hsize
\csname@twocolumnfalse%
\endcsname
\draft
\title{Modulational instability in Bose-Einstein
condensates in optical lattices}
\author{V. V. Konotop$^\dag$ and M. Salerno$^\ddag$}
\address{$^\dag$Departmento de F\'{\i}sica and Centro de F\'{\i}sica da Mat\'eria
         Condensada, Universidade de Lisboa, \\
         Complexo Interdisciplinar, Av. Prof. Gama Pinto 2, Lisboa 1649-003,
         Portugal \\
         $^\ddag$Dipartimento di Fisica "E.R. Caianiello",
         Universit\'a di Salerno, I-84081 Baronissi (SA), Italy, and \\
         Istituto Nazionale di Fisica della Materia (INFM), Unit\'a
         di
         Salerno, Italy.}
\maketitle
\begin{abstract}
A self consistent theory of a cylindrically shaped Bose-Einstein
condensate (BEC) periodically modulated by a laser beam is
presented. We show, both analytically and numerically, that
modulational instability/stability is the mechanism by which
wavefunctions of soliton type can be generated in cylindrically shaped BEC subject to a one-dimensional optical lattice. The theory explains why bright
solitons can exist in BEC with positive scattering length and why
condensate with negative scattering length can be stable and give
rise to dark solitary pulses. PACS numbers: {42.65.-k, 42.50.
Ar,42.81.Dp\\}
\end{abstract}
 ]

There is an increasing interest in the study of Bose-Einstein
condensates (BEC) in presence of periodic potentials, such as the
one induced by detuned standing waves of light (optical lattices)
\cite{berg}. Switching on an optical lattice in a continuous
BEC induces fragmentation of the original wavefunction into
local wavefunctions centered around the minima of the potential,
this leading to a crystal-like structure of mutually interacting
BECs. In analogy with the usual theory of crystals, one can think
to control the dynamics of this new state of matter by properly
choosing the parameters of the lattice. This gives, for example,
the possibility to observe macroscopic quantum interference
phenomena with emission of coherent pulses of atoms (Bloch
oscillations), as recently reported in Ref. \cite{Kasevich} for
vertical BEC arrays in the gravitational field. Understanding the
properties of BEC in optical lattices is therefore of fundamental
importance for developing novel applications of quantum mechanics
such as atom lasers and atom interferometers. For small
overlapping between local wavefunctions, a tight-binding model can
be developed. This was done, for the one dimensional (1D) case, in
Refs. \cite{tromb} where it was shown that the mean field equation
for the condensate wavefunction reduces to the so called discrete
nonlinear Schr\"odinger equation \cite{scott}.
The tight-binding approximation, however, putting restrictions on
the shape of the wave function (i.e. on the number of atoms in the
condensate), as well as on the potential profile, is applicable
only to particular experimental settings.
From this point of view it is desirable to develop a theory of BEC
in optical lattices which does not rely on this approximation.
Studies in this direction were made in terms of a 1D nonlinear
Schr\"odinger equation (NLS) with trigonometric \cite{Niu} or
elliptical potentials \cite{bronski}. Bright and dark solitons in
BEC in optical lattices, analogue to the gap-soliton of photonic
crystals \cite{sterke}, were also shown to exist
\cite{steel,potting}.

The aim of this Letter is to investigate, both analytically and
numerically, modulational instability phenomena of extended states
at the border of the Brillouin zone. To this end we construct
approximate ground state solutions of the original 3D by means of
a multiple scale expansion,  starting from the exact
eigenfunctions of the underlying linear Schr\"odinger equation
with potentials which are parabolic in the transverse direction
and periodic in the longitudinal one (periodic cylindrical trap).
We show that at the lowest orders in the expansion the condensate
evolves according to an effective 1D NLS with the dispersive term
depending on the effective mass of the Bloch states of the
underlying linear problem. Extended states close to the borders of
the Brillouin zone, are then shown to be unstable (stable) against
small spatial modulations (modulational instability) depending on
the sign of the dispersion in the effective 1D NLS. The stability
properties of these states is shown to be the basic mechanism by
which bright (resp. dark) solitons are created in BEC with
positive (resp. negative) scattering lengths. Numerical
simulations of the longitudinal BEC dynamics confirm the
predictions of our theory. The possibility to observe the
modulational instability phenomena in real BEC is discussed
at the end of the paper.

As is well known \cite{gpe}, the condensate wavefunction is
described by the Gross-Pitaevskii equation (GPE)
\begin{equation}
\label{GPE0} i\hbar \frac{\partial \Psi({\bf r},t)}{\partial t} =
[-\frac{\hbar^2}{2m}\Delta + V({\bf r})+g_0[\Psi({\bf
r},t)[^2]\Psi({\bf r},t)
\end{equation}
with $g_0 = 4\pi \hbar^2 a_s /m$, $m$ is the atomic mass, and
$a_s$ is the $s$-wave scattering length of atoms which can be
either positive or negative. We consider a trap potential of the
form $V(\br)=\frac 12 m \nu^2 \brt^2 +V_0\cos(\kappa z)$, which
model a cylindrically shaped BEC periodically modulated along the $z$-axis
(the results, however, will not depend on the form of periodic
potential used, and can be easily generalized to arbitrary
$z$-periodic potentials). Here $\br\equiv(\brt,z)$, $V_0$ is the
potential deepness, $\nu$ the trap frequency in the transverse
direction, and $2\pi/\kappa$ the period of the modulation. We
assume periodic boundary conditions $\Psi(\brt,z,t)=\Psi(\brt,
z+L,t)$, with $L$ denoting the length of the cylinder. The change of
variables $t\mapsto 2t/\nu$, $\br\mapsto a_0 \br$, $\Psi\mapsto
(N/a_0^3)^{1/2}\psi$, with $a_0=(\hbar/(m\nu))^{1/2}$, allows us to
rewrite Eq.(\ref{GPE0}) in the dimensionless form
\begin{equation}
\label{GPE} i\frac{\partial \psi({\bf r},t)}{\partial t} = [ {\cal
L} +\chi [\psi({\bf r},t)[^2]\psi({\bf r},t),
\end{equation}
where $\chi=8\pi Na_s/a_0$, and ${\cal L}\equiv \lt+\lz$ with
\begin{eqnarray}\label{linop}
{\cal L}_{\bot}=-\Delta_\bot+\rt^2, \, \qquad {\cal
L}_{z}=-\partial^2/\partial z^2+2\Lambda \cos(kz),
\end{eqnarray}
(here $\Delta_\bot$ denotes  the two-dimensional Laplacian, $k=a_0/\kappa$ and
$\Lambda=V_0\nu/\hbar$). In these units the wave function
results normalized to one, i.e.
\begin{equation}
\label{norm}
\int d\brt\int_{0}^{\tilde{L}}dz [\psi[^2 = 1,
\end{equation}
with $\tilde{L} \equiv L/a_0$ denoting the normalized length of
the cylinder. In the following we shall restrict to the small
amplitude limit ($\chi |\psi|^2 \ll 1$)  and construct a solution
of Eq. (\ref{GPE}) perturbatively, starting from the solution of
the linear problem. These last can be written as products of
eigenfunctions of the operators in Eq. (\ref{linop})
\[
\lz \phi_{\tilde{n}q}(z)={\cal
E}_{\tilde{n}q}\phi_{\tilde{n}q}(z),
\,\,\, \lt \xi_{n m}(\brt)=\varepsilon_{n m}\xi_{n m}(\brt) .
\]
For the considered potential,  $\phi_{\tilde{n}q} (z)$ are
solutions of the Mathieu equation, while $\xi_{nm}(\brt)$ are
eigenfunctions of the two-dimensional harmonic oscillator ($n$ and
$m$ denote the principal and the angular quantum numbers of the
harmonic oscillator, while $\tilde{n}$ and $q$ denote the band
index and  the wavevector inside the first Brillouin zone of the
1D lattice, respectively). We look for solutions of Eq.
(\ref{GPE}) of the form
\begin{equation}
\label{expans} \psi=\sqrt{\frac{\tilde
{L}}{[\chi[}}\,\,\,(\sigma\psi_1+\sigma^2\psi_2+\cdots),
\end{equation}
with $\sigma$ a small parameter whose physical meaning will be
clarified later (the prefactor is unimportant and introduced just
for convenience). Since we are interested in the ground state we
take as the leading order term in Eq. (\ref{expans}) a small
modulation of the linear ground state wavefunction ($n_0=0$,
$m_0=0$, $\tilde n_0=1$) of the form
\begin{equation}
\label{zorder}
\psi_1=A({\bf z},\,{\bf t})\phi_{\tilde{n}_0q}
(z_0)\xi_{n_0m_0}(\brt) e^{-i\omega_{n_0m_0\tilde{n}_0}(q) t_0},
\end{equation}
with $\omega_{n_0,m_0,\tilde{n}_0}(q)=\varepsilon_{n_0,m_0}+ {\cal
E}_{\tilde{n}_0q}\equiv\omega (q)$. The modulating amplitude
$A(\bf{z},\,\bf{t})$ is considered to be a function of a set of
independent spatial and temporal variables of the form ${\bf
z}\equiv (z_1, z_2, \dots , z_n, \dots)$ with $z_n=\sigma^n z$,
and ${\bf t}\equiv (t_1, t_2, \dots , t_n, \dots)$ with
$t_n=\sigma^n t$, respectively. To simplify the notation we
introduce the shortcut symbols $\varepsilon_{0} \equiv
\varepsilon_{n_0,m_0}$, ${\cal E}(q) \equiv {\cal
E}_{\tilde{n}_0,q} $, $\phi_{q}(z) \equiv
\phi_{\tilde{n}_0,q}(z)$, and in the modulation amplitude $A$, we
show only the dependence on the most ``rapid" variables. The time
and coordinate derivatives in Eq. (\ref{GPE}) are then expanded as
${\partial}/{\partial t}=\sum_{\alpha=0}
\sigma^{\alpha}{\partial}/{\partial t_{\alpha}}$ and
${\partial}/{\partial z}=\sum_{\alpha=0} \sigma^{\alpha}
{\partial}/{\partial z_{\alpha}}$.  Substituting the above
expansions in Eq. (\ref{GPE}) and collecting all the terms of the
same order in $\sigma$, we obtain at the first order:
$i{\partial\psi_1}/{\partial t_0}-{\cal L}\psi_1=0 $, which is
evidently satisfied by $\psi_1$ given by (\ref{zorder}). At the
second order in $\sigma$, the following equation is obtained
\begin{equation}
\label{firstorder} i\frac{\partial\psi_2}{\partial t_0}-{\cal
L}\psi_2 = -i\frac{\partial\psi_1}{\partial
t_1}-2\frac{\partial^2\psi_1} {\partial z_0\partial z_1},
\end{equation}
whose solution can be searched in the form
\begin{equation}
\label{forder00} \psi_2=\sum_{n,m}\sum_{(\tilde{n},q')\neq (\tilde{n}_0,q)}   B_{n,m,\tilde{n}}(q')\phi_{\tilde{n}q'}
\xi_{nm}e^{-i\omega(q)t_0}.
\end{equation}
Substituting (\ref{forder00}) in (\ref{firstorder}) and projecting
along the eigenfunctions of operators (\ref{linop}) with
$\tilde{n}\neq\tilde{n}_0$, we find that
\begin{equation}
\label{forder0} \psi_2=\frac{\partial A}{\partial
z_1}\sum_{\tilde{n}\neq \tilde{n}_0}
\frac{\Gamma_{\tilde{n}\tilde{n}_0}}{\omega_{0}(q)- \omega_{n_0
m_0\tilde{n}}(q)}\phi_{\tilde{n}q} \xi_{n_0m_0}e^{-i\omega(q)t_0}
\end{equation}
with $\Gamma_{\tilde{n}\tilde{n}_0}(q)=-2\int_{0}^{\tilde{L}}
\bar{\phi}_{\tilde{n}q}(z)\frac{d}{dz}\phi_{\tilde{n}_0q}(z)\,dz.$
%
The solvability condition of Eq. (\ref{firstorder}) reads as
$\frac{\partial A}{\partial t_1}+v\frac{\partial A}{\partial
z_1}=0$, from which we see that $A\equiv A(\zeta;z_2,t_2)$, with
$\zeta = z_1-v t_1$. Note that $ v\equiv
v(q)=i\Gamma_{\tilde{n}_0\tilde{n}_0}(q) $ can be interpreted as
the group velocity of the wave packet in the $z$-direction.
Finally, at the third order in $\sigma$, we get
\begin{eqnarray}
\label{secorder} i\frac{\partial\psi_3}{\partial t_0}-{\cal
L}\psi_3&=& -i\frac{\partial\psi_1}{\partial
t_2}-i\frac{\partial\psi_2} {\partial t_1}
-2\frac{\partial^2\psi_2}{\partial z_0\partial z_1} \nonumber
\\ &-&\left(\frac{\partial^2}{\partial z_1^2}+
2\frac{\partial^2}{\partial z_0\partial z_2}\right)\psi_1+ \chi
[\psi_1[^2\psi_1.
\end{eqnarray}
Requiring orthogonality (to avoid secular terms) between  the
right hand side of this equation and the kernel of the operator
$i\partial/\partial t_0-{\cal L}$, and taking into account the
expressions of $\psi_1$ and $\psi_2$ derived above, we find  that
Eq.(\ref{secorder}) reduces to the following NLS equation
\begin{equation}
\label{nls1} -i\left(\frac{\partial A }{\partial
t_2}+v\frac{\partial A}{\partial z_2} \right)-D\frac{\partial^2
A}{\partial \zeta^2}+\tilde{\chi}[A[^2 A=0,
\end{equation}
where $D\equiv D(q)=1+\sum_{\tilde{n}\neq \tilde{n}_0}
\frac{|\Gamma_{\tilde{n}\tilde{n}_0}(q)|^2}{\omega(q)-\omega_{n_0m_0\tilde{n}}(q)}
$
is the effective group velocity dispersion induced by the periodic
potential, and
\begin{eqnarray}
\label{chi} \tilde{\chi}
=\mbox{sign}\,(\chi)\frac{\tilde{L}}{2\pi}  \int_{0}^{\tilde{L}}
[\phi_{q}(z)[^4\,dz
\end{eqnarray}
is the effective nonlinearity (here we integrated on radial
variables and used the ground state wavefunction of the 2D
harmonic oscillator). The above expressions of $v$ and $D$, in
terms of eigenfunctions of the linear operator ${\cal L}$, can be
simplified by expressing them in terms of the energy spectrum of
the non interacting linear system. This can be done in the same
manner as in the theory of optical gap solitons \cite{sterke}.
To this end, we take two close Bloch
solutions of the 1D linear problem, of the form $\phi_q(z)=\exp (i
q z)u_{\tilde n, q}(z)$, which differ only by a small $\delta q$,
so that $u_{\tilde{n},q+\delta q}(z)$ can be considered as a
perturbation of $u_{\tilde{n},q}(z)$ generated by the operator
$-2i\delta q\left(\frac{d}{dz}+iq\right)+(\delta q)^2$. This
perturbation produces a shift $\Delta={\cal E}_{\tilde{n},
q+\delta q}-{\cal E}_{\tilde{n}, q}$ in energy, which can be
expanded in a Taylor series in $\delta q$. On the other hand,
$\Delta$ can also be computed from perturbation theory. A
comparison of the corresponding expressions leads to
$v={d\omega(q)}/{dq}$ and $D=\frac{1}{2}
\frac{d^2\omega(q)}{dq^2}$, i.e. $v$ and $D$ are, respectively,
the slope (velocity) and the  curvature (inverse effective mass)
of the energy band (Bloch states) of the underlying linear
problem.

From the physical point of view the above results  have a number of
consequences. First,  the group velocity induced by the
periodicity at the boundaries of zone dominate the dispersion
inherent to NLS. For example, if we take $k=2.0$ and
$\Lambda=0.5$, we have that the edges of the first gap $[{\cal
E}^{(1)}, {\cal E}^{(2)}]$ are at ${\cal E}^{(1)}\approx 0.47$,
and ${\cal E}^{(2)}\approx 1.47$. The effective dispersion at
these points is $\omega_1'' \approx -6.13$, and $\omega_2''\approx
10.14$, respectively (here $\omega_j''=d^2\omega/dq^2|_{q=q_j}$).
Thus even in the case the group velocity dispersion does not
change sign it becomes much larger than the NLS dispersion.
Second, for fixed nonlinearity and in presence of the periodic
potential, the dynamics will crucially depend on the sign of $D$.
This sign can be controlled by changing the wave number of the
initial state, as well as, the potential parameters. Instability
phenomena of extended (Bloch) states close to the edges of the
Brillouin zone can then appear. To understand this, let us assume
positive scattering length ($\tilde \chi>0$ in (\ref{nls1})) and
consider the Bloch state at ${\cal E}^{(1)}$, for which
$D^{(1)}<0$. In the presence of a repulsive inter-atomic
interaction ($\chi>0$), the energy of this state will be shifted
upward in the gap where it cannot exists. One can expect then the
state to become unstable against small spatial modulations
(modulational instability) so that new excitations must arise.
Equation (\ref{nls1}) predicts that out of the instability bright
solitons should appear (recall that for $\tilde\chi>0$, and $D<0$
(resp.$D>0$), Eq. (\ref{nls1}) has stable bright (resp. dark)
soliton solutions).
On the contrary, if we take as initial state the Bloch state at
the bottom of the second band, ${\cal E}^{(2)}$, where
$D=D^{(2)}>0$, one expects modulational stability instead (in this
case the nonlinearity is pulling the energy of the state further
up in the second band where it can still exist). This extended
stable state can be then used as background to construct the dark
soliton solution expected in this  case from Eq.(\ref{nls1}) (see
below). Obviously, for negative scattering lengths the opposite
situation will occur i.e. modulational instability will appear at
the top of the gap (leading to bright solitons) and stability at
the bottom (leading to dark solitons). From this it is clear that
the stability properties of the Bloch states at the edge of the
Brillouin zone, plays a crucial role for the existence of bright
and dark solitons in BEC in optical lattices both for positive and
negative scattering lengths.

These predictions can be easily checked by direct numerical
integration. To this regard we remark that instabilities along
$z$-direction mainly depend on the spectrum of the operator $\lz$
(the transverse distribution of the condensate affects only the
absolute value of the coefficient $\tilde\chi$), so that we can
perform numerical simulations in the framework of a 1D NLS
equation obtained from Eq. (\ref{GPE}) with ${\cal L}\approx{\cal
L}_{z}$. Moreover, we note that the Bloch state (Mathieu function)
at the top of the first band (bottom of the gap), is an odd
function of $z$ which can be approximated by $\sin(z)$, while the
one at the bottom of the second band (top of the gap) is an even
function of $z$ very close to $\cos(z)$. In the following we shall
use these approximate states as initial conditions for
investigating modulational stability since they are, in real
experiments, more easy to generate.
\begin{figure}[h]
\caption{Modulational instability in Eq. (\ref{GPE}) with
${\cal L}\approx{\cal L}_z$ for parameter values
$\tilde{\chi}=1.0$,  $k=2.0$ and $\Lambda=0.5$. The initial
condition is an approximated eigenfunction, taken as a sine
function, of the first band of the linear system at the edge
${\cal E}^{(1)}\approx 0.47$ of the Brillouin zone.}
\label{figone}
\end{figure}
In Fig.~\ref{figone} a numerical simulation of the 1D problem with
initial condition close to the state at the bottom of the gap, is
depicted. We see that, as expected from our analysis, modulational
instability develops and, in spite of the fact that we have
positive scattering ($\tilde{\chi} = 1$), bright solitons are
created in agreement with our analysis (the number of solitons
coming out from the instability can be estimated as $L\,k_{max}/(2
\pi)$, where $k_{max}$ is the wavenumber of the most unstable
linear mode \cite{ruffo}). We remark that although the theory is
valid for small amplitude excitations, the numerical simulations
show that the obtained results extend also above this limit (note
that in Fig.~\ref{figone} $\tilde\chi=1$). An intuitive
explanation for this is that small amplitude solitons once formed
can only become more and more localized as the nonlinearity is
increased. The modulational instability at higher nonlinearity
should therefore produce solitons which are more localized and of
large amplitude. This is precisely what observed in Fig 1. In
contrast to this, we find that an initial condition corresponding
to a Bloch state close to the top of the gap, remains
modulationally stable also in presence of nonlinearity. This is
reported in Fig.~\ref{figtwo} for an initial profile of cosine
type. It is interesting to note that one can use this state to
construct the stable dark soliton predicted by Eq. (\ref{nls1}).
To this end we take as initial condition a modulated Bloch state
of the form $\tanh(\lambda z) \cos(z)$, where the cosine function,
taken as background, approximates the Mathieu eigenfunction at the
edge ${\cal E}^{(2)}$, while the $\tanh$ modulation is used to
make the profile close to the expected dark state.
\begin{figure}[h]
\caption{Same as in Fig.1 but for the
eigenfunction at the top of the gap ${\cal E}^{(2)}\approx 1.47$
approximated with a cosine function.} \label{figtwo}
\end{figure}
\begin{figure}[h]
\caption{Same as in Fig.1 but for a dark soliton initial
condition} \label{figthree}
\end{figure}
In Fig.~\ref{figthree} the corresponding numerical simulation is
reported, from which we see that a dark soliton is indeed
generated, in perfect agreement with our analysis (the energy of
this state is in the gap close to the bottom of the second band).
It is interesting to note that for negative scattering lengths
this leads to the existence of dark soliton in BEC in optical
lattices (in this case one must use the stable Bloch state at the
bottom of the gap as background for the dark solution).

In order to check the self-consistency of the theory  we shall
estimate the size of the parameter $\sigma$ used for the
expansion, and the magnitude of the effective nonlinearity in Eq.
(\ref{nls1}). To this end we start with the dark soliton or
periodic solution and notice that the eigenfunctions $\phi_q(z)$
are normalized to one, so that $\tilde{L}|\phi_q|^2\sim 1$ and
hence, from Eq. (\ref{chi}) we have that $\tilde{\chi}\sim 1$.
Similarly, from the normalization of the wavefunction (\ref{norm})
and from the expansion (\ref{expans}), we have that ${\pi
\tilde{L}^2}\sigma^2/{|\chi|}\sim 1$ from which, after restoring
physical units, we get $\sigma^2=8 N {a_s a_0}/{L^2}$.
If we consider the case of a condensate with $N\approx 10^4$ atoms
of $^{87}Rb$ ($a_s\sim 5.5$nm) with a radial size $a_0\sim
17\,\mu$m, and length $L\sim 300\,\mu$m
we have that $\sigma^2 \sim 0.08$, this being reasonably small to
justify our expansion (smaller values can be achieved by
considering longer, thinner, cylinders and smaller values of N).
As to the initial state, we remark that it could be generated from
an uniform cylinder by modulating it along the z-axis with a sine
or cosine wave of light with twice the wavelength of the lattice.
Another possibility is to use an initial uniform condensate and
accelerate the lattice until the state reaches the edge of the
band (it is enough to be close to the edge for the instability to
develop). Finally we mention, that although the expansion has been
provided for a cylindrically-shaped BEC, a number of effects
discussed is relevant to a cigar-shaped BEC (i.e. including
parabolic confining potential in the direction of periodicity).
This is the case when the effect (instability, bright or static
dark soliton) observed has a scale much less than the length of
the condensate. We hope that the phenomena of modulational
instability discussed in this Letter will be soon observed in real
BEC experiments.

\vskip .15cm
VVK acknowledges support from FEDER, the Program
PRAXIS XXI, grant N$^o$ /P/Fis/10279/1998, and the European grant,
COSYC n.o HPRN-CT-2000-00158. MS thanks the "Centro de F\'{\i}sica
da Materia Condensada", University of Lisbon, for a one month
Visiting Professorship, and the MURST for financial support
through the PRIN-2000 Initiative.

\end{document}